\documentclass[twocolumn,showpacs,prb]{revtex4}
\usepackage{amsmath}
\usepackage{amsfonts}
\usepackage{graphicx}
\usepackage{color}
\usepackage{hyperref}

\renewcommand{\Re}{\mathop\mathrm{Re}\nolimits}
\renewcommand{\Im}{\mathop\mathrm{Im}\nolimits}
\newcommand{\tr}{\mathop{\rm tr}\nolimits}
\newcommand{\sgn}{\mathop{\rm sign}\nolimits}

\begin{document}

\title{Anisotropic conductivity tensor
on a half-filled high Landau level}
\author{I.S. Burmistrov}
\affiliation{Landau Institute for Theoretical Physics, Kosygina
str. 2, 117940 Moscow, Russia} \affiliation{Institute for
Theoretical Physics, University of Amsterdam, Valckenierstraat 65,
1018XE Amsterdam, The Netherlands}

\begin{abstract}
We study two-dimensional interacting electrons in a weak
perpendicular magnetic field with the filling factor $\nu \gg 1$
and in the presence of a quenched disorder. As it is known, the
unidirectional charge density wave state can exist near a
half-filled high Landau level at low temperatures if disorder is
weak enough. We show that the existence of the unidirectional
charge density wave state at temperature $T<T_c$ where $T_c$ is
the transition temperature leads to the anisotropic conductivity
tensor. We find that the anisotropic part of conductivity tensor
is proportional to $(T_c-T)/T_c$ below the transition in
accordance with the experimental findings. The order parameter
fluctuations wash out the mean-field cusp at $T=T_c$ and the
conductivity tensor becomes anisotropic even above the mean-field
transition temperature $T_c$.
\end{abstract}

\maketitle

\section{Introduction}

A two-dimensional electron gas (\textit{2DEG}) in a perpendicular
magnetic field ($H$) have remained a subject of intensive studies,
both theoretical and experimental, for several decades~\cite{AFS}.
Recently the phenomenon of sharp anisotropy of magnetoresistance
near half-filled high Landau levels with Landau level index $N =
2,3,4$ at low temperatures have been
discovered~\cite{LCEPW,DTSPW}. Right away, the magnetoresistance
anisotropy have been related with the possible existence of
unidirectional charge density wave (\textit{UCDW}) state near
half-filling of a high Landau level that theoretical results of
Refs.~\cite{KFS,FKS,MC} had been predicted in the limit $N\gg 1$.
Later a bunch of theoretical and experimental papers have
appeared~\cite{Fogler}. However, only a few theoretical researches
have been concerned the transport properties of the system whereas
the only experimentally measured quantity is magnetoresistance.

In the limit of zero temperature the unidirectional charge density
wave has strongly non-sinusoidal profile with well-defined
edges~\cite{KFS} and the conductivity tensor can be evaluated in
the framework of an effective theory for edge
excitations~\cite{MF}. The result of Ref.~\cite{MF} is in a
qualitative agreement with measured magnetoresistance behavior in
the limit of zero temperature~\cite{DTSPW,ELCPW}. Later the
``semicircle law'' of Dykhne-Ruzin~\cite{DR} have also been
extended to the case of the anisotropic conductivity
tensor~\cite{vOHS}.

At present a thorough analysis of temperature ($T$) dependence of
magnetoresistance near the transition temperature $T_c$ from the
liquid state to the UCDW state is absent. The main objective of
the paper is to present such analysis for the domain $|T_c-T|/T_c
\ll 1$ where the expansion in the UCDW order parameter is
legitimate. The effect of the order parameter fluctuations that
are enhanced near the phase transition on the conductivity tensor
both below and above the transition temperature $T_c$ is also
investigated.

We assume the presence of a \textit{weak} random potential created
by impurities near 2DEG such that the Landau level broadening
$1/\tau$ is much less the spacing $\omega_H$ between Landau
levels, $1/\tau\ll\omega _{H}$. Here the $\omega_H$ is cyclotron
frequency $\omega_H=e H/m$ with $e$ and $m$ being the electron
charge and band mass respectively (we use the units with $\hbar
=1$, $c=1$, and $k_{B}=1$).

One among the main results of the paper is the fact that the
conductivity tensor $\sigma_{ab}$ (we measure conductivity in
units of $e^2/h$) of two-dimensional electrons in the UCDW state
acquires anisotropic part $\sigma_{ab}^{(\rm anis)}$ proportional
to temperature deviation from $T_c$, i.e. $\sigma_{ab}^{(\rm
anis)} \propto (T_c-T)/T_c$. Other main result is that in the
vicinity of the transition temperature $T_c$ there is additional
anisotropic contribution $\delta\sigma_{ab}^{(\rm anis-f)}\propto
(|T_c-T|/T_c)^{-3/2}$ that we will refer as fluctuational to the
conductivity tensor $\sigma_{ab}^{(0)}$ of the liquid state due to
UCDW order parameter fluctuations.

We start out with an introduction to the formalism that mainly
follows one introduced in the previous paper~\cite{Burm2}. In
section~\ref{ch3.1} the effective action of ``three level model''
is developed. The conductivity tensor is evaluated in
section~\ref{ch3.2}. The effect of order parameter fluctuations is
investigated in section~\ref{ch3.3}. In section~\ref{ch3.4} the
results obtained are discussed in relation with recent experiments
~\cite{LCEPW,DTSPW}.

Some of the results of the present paper have been published in a
brief form in Ref.~\cite{Burm3}.

%
%
\section{\label{ch3.1}``Three level'' model}

%
\subsection{\label{ch3.1.1}Introduction}

To start out consider the system of two-dimensional interacting
electrons in the presence of a random potential $V_{\rm
dis}(\textbf{r})$ and perpendicular magnetic field $H$. The
parameter that characterizes the strength of the Coulomb interaction is $r_{s}=\sqrt{2}%
e^{2}/\varepsilon v_{F}$ with $v_{F}$ being the Fermi velocity and $%
\varepsilon$ the dielectric constant of a media. We assume that
the Coulomb interaction between the electrons is weak, $r_{s}\ll
1$, and the magnetic field obeys the condition $Nr_{s}\gg 1$ where
the Landau level index $N=[\nu /2]$ is the integer part of half of
filling factor $\nu$. In addition, we assume that the Landau level
broadening $1/\tau$ is not too small, $1/\tau\gg \omega_H N^{-1}
\ln \sqrt{2}r_s N$. In this case it is possible to construct an
effective field theory for electrons on the highest partially
filled Landau level by integrating out all other degrees of
freedom~\cite{AG,Burm1}.

Also, we consider the case when the Landau levels are
spin-resolved that occurs according to Ref.~\cite{AG} if $1/\tau
\ll \Delta_{\rm ex} = (r_s \omega_H /\pi\sqrt{2})\ln
2\sqrt{2}/r_s$. Therefore, the Landau level broadening should be
restricted from below and from above as
\begin{equation}\label{eq000}
\frac{\omega_H}{N} \ln \sqrt{2}r_s N \ll \frac{1}{\tau} \ll
\frac{r_s \omega_H}{\pi\sqrt{2}}\ln\frac{2\sqrt{2}}{r_s}.
\end{equation}

Throughout the paper we characterize the UCDW state by the order
parameter $\Delta (\mathop{\bf Q})$. All calculations are
performed under the assumption $Nr_{s}^{2}\gg 1$. In this case the
Hartree-Fock approximation is well justified~\cite{MC} moreover it
is known that the corrections to it are small in the parameter
$a_{B}/l_{H}=1/Nr_{s}^{2}\ll 1$, where $a_{B}=\varepsilon /m
e^{2}$ is the Bohr radius and $l_{H}=1/\sqrt{m\omega _{H}}$ the
magnetic length.

\subsection{\label{ch3.1.0}Formalism}

Thermodynamic potential of the system in hand can be written as
\begin{equation}
\Omega =-\frac{T}{N_{r}}\int \mathcal{D}[\overline{\psi },\psi
]\int \mathcal{D}[V_{\rm dis}]\,\mathcal{P}[V_{\rm dis}]\,\exp
\mathcal{S}[\overline{\psi },\psi ,V_{\rm dis}].  \label{ZStart}
\end{equation}
where action $\mathcal{S}[\overline{\psi },\psi ,V_{\rm dis}]$ is
written in Matsubara representation
\begin{gather}
\mathcal{S}=\int d\textbf{r}\sum\limits_{\alpha,\omega_n,\omega_m}
\overline{\psi _{\omega_{n}}^{\alpha,\sigma}}(\textbf{r})
\Bigl[(i\omega_{n}+\mu -V_{\rm dis}(\textbf{r}))\delta_{n m}
-\mathcal{\hat H}\Bigr] \notag \\ \times \psi
_{\omega_{m}}^{\alpha,\sigma }(\textbf{r}) -
\frac{T}{2}\sum_{\omega_{n},\omega_{m}, \nu_{l}}\int
d\textbf{r}d\textbf{r}^{\prime}\overline{ \psi
_{\omega_{n}}^{\alpha,\sigma }}(\textbf{r}) \psi
_{\omega_{n}-\nu_{l}}^{\alpha,\sigma }(\textbf{r}) \notag\\ \times
U_0(\textbf{r}-\textbf{r}^{\prime})\overline{\psi
_{\omega_{m}}^{\alpha,\sigma^{'} }}(\textbf{r})\psi _{\omega_{m}
+\nu_{l}}^{\alpha,\sigma^{'} }(\textbf{r}^{\prime }).
\label{Sinitc3}
\end{gather}
Here, $\psi _{\omega_{n}}^{\alpha,\sigma}(\textbf{r})$ and
$\overline{\psi _{\omega_{n}}^{\alpha,\sigma}}(\textbf{r})$ are
annihilation and creation electron operators. $T$ stands for
temperature, $\mu$ chemical potential, $\sigma$ and $\sigma^{'}$
spin indices, $\omega_{n}=\pi T(2n+1)$ fermionic frequency whereas
$\nu_{n}
 =2\pi Tn$ bosonic one. Matrix $\mathcal{\hat H}$ is defined as
\begin{equation}\label{eq5}
\mathcal{\hat H} = \sum_{\alpha \nu_n} \mathcal{H}(\nu_n)
I^{\alpha}_{n},
\end{equation}
with matrices
\begin{equation}\label{eq6} (I^{\alpha}_{n})^{\beta
\gamma}_{k l} = \delta^{\alpha \beta} \delta^{\alpha \gamma}
\delta_{k-l,n}
\end{equation}
being $U(1)$ generators. One-particle hamiltonian $\mathcal{H}$
describes a two-dimensional electron in constant perpendicular
magnetic field $H=\epsilon _{ab}\partial _{a}A_{b}$ and in
time-dependent magnetic field with vector-potential $\textbf{a}$,
\begin{equation}\label{H0c3}
\mathcal{H}=\frac{1}{2m}(-i\nabla -e\textbf{A}- e\,
\textbf{a})^{2}.
\end{equation}

As usual, we assume the white-noise distribution for the random
potential,
\begin{equation}
\mathcal{P}[V_{\rm dis}(\textbf{r})]=\frac{1}{\sqrt{2\pi g}}\exp
\left( -\frac{1}{2 g} \int d\textbf{r} V_{\rm dis}^2(\textbf{r})
\right). \label{disc3}
\end{equation}
In order to average over disorder we introduce $N_r$ replicated
copies of the system~\cite{EA} labelled by the replica indices
$\alpha=1, \cdots, N_r$.

It is convenient to rewrite one-particle hamiltonian \eqref{H0c3}
with a help of covariant derivative
\begin{equation}\label{covc3}
\textbf{D} = \nabla- i e \textbf{A},
\end{equation}
in order to extract the time-dependent vector potential
$\textbf{a}(\nu_n)$,
\begin{gather}\label{H1c3}
\mathcal{H} = -\frac{1}{2m}\textbf{D}^2 + K(\nu_n), \\
K(\nu_n) = -\frac{e}{m}\textbf{a}(\nu_n) \textbf{D}
+\frac{e^2}{2m} \sum_{\nu_m} \textbf{a}(\nu_{n-m}) \textbf{
a}(\nu_m).\notag
\end{gather}

%
\subsection{\label{ch3.1.2}Effective action of ``three-level'' model}
%

To investigate the thermodynamic properties of electrons on the
$N$th Landau level one can integrate out electrons on all other
Landau levels~\cite{AG}. However, to find conductivity tensor
projection on the single $N$th Landau level is not appropriate
because of covariant derivative $\textbf{D}$ has non-zero matrix
elements only for transitions between adjacent Landau levels. It
is necessary therefore to consider not only the $N$th Landau level
alone but two adjacent ones, the $(N-1)$th and $(N+1)$th Landau
levels.

Extending the projection to the $N$th Landau level only of
Refs.~\cite{AG,Burm1}, we obtain effective action for electrons on
the $(N-1)$th, $N$th, and $(N+1)$th Landau level as follows
\begin{gather}
\mathcal{S}=\int d\textbf{r}\sum\limits_{\alpha,\omega_n,\omega_m}
\overline{\psi _{\omega_{n}}^{\alpha,\sigma}}(\textbf{r})
\Bigl[(i\omega_{n}+\mu - V_{\rm dis}(\textbf{r}))\delta_{n m}
-\mathcal{\hat H}\Bigr]\notag \\\times \psi
_{\omega_{m}}^{\alpha,\sigma}(\textbf{r}) -
\frac{T}{2}\sum_{\omega_{n},\omega_{m}, \nu_{l}}\int
d\textbf{r}d\textbf{r}^{\prime}\overline{ \psi
_{\omega_{n}}^{\alpha,\sigma}}(\textbf{r}) \psi
_{\omega_{n}-\nu_{l}}^{\alpha,\sigma}(\textbf{r})\notag \\
\times U_{\rm scr} (\textbf{r}-\textbf{r}^{\prime}) \overline{\psi
_{\omega_{m}}^{\alpha,\sigma^{'}}}(\textbf{r})\psi _{\omega_{m}
+\nu_{l}}^{\alpha,\sigma^{'}}(\textbf{r}^{\prime }).\label{S3Lc3}
\end{gather}
Here $\psi _{\omega_{n}}^{\alpha,\sigma}(\textbf{r})$ and
$\overline{\psi _{\omega_{n}}^{\alpha,\sigma}}(\textbf{r})$ are
annihilation and creation operators of an electron on the
$(N-1)$th, $N$th, and $(N+1)$th Landau levels,
\begin{equation}\label{PsiDagc3}
\psi^{\alpha,\sigma}_{\omega_n}(\textbf{r}) =
\sum\limits_{p=N-1}^{N+1}\psi_{p
\omega_n}^{\alpha,\sigma}(\textbf{r}),\qquad
\overline{\psi_{\omega_n}^{\alpha,\sigma}} (\textbf{r}) =
\sum\limits_{p=N-1}^{N+1}\overline{\psi^{\alpha,\sigma}_{p\omega_n}}(\textbf{r}).
\end{equation}
The screened electron-electron interaction $U_{\rm
scr}(\textbf{r})$ has the following Fourier transform
\begin{eqnarray}
U_{\rm scr}(q) &=& \frac{2\pi e^{2}}{\varepsilon q}\Biggl [
1+\frac{2}{
qa_{B}}\left ( 1-\frac{\pi }{6\omega _{H}\tau }\right )\label{U01c3} \\
&\times & \left( 1-\mathcal{J} _{0}^{2}(qR_{c})-2
\mathcal{J}_{1}^{2}(qR_{c})\right) \Biggr ]^{-1}.\notag
\end{eqnarray}
It is different from one obtained in Ref.~\cite{Burm1,Foot1}. The
reason for that is exclusion of contributions from the $(N-1)$th
and $(N+1)$th Landau level from the polarization operator.

Effective action \eqref{S3Lc3} was obtained under assumptions
discussed in Sec.~\ref{ch3.1.1}. Hereafter, for reasons to be
explained shortly we neglect small correction $\pi/(6\omega
_{H}\tau)\ll 1$ in the screened electron-electron interaction
~\eqref{U01c3}.

%
%
\subsection{\label{ch3.1.3}Hartree-Fock decoupling}

Effective action ~\eqref{S3Lc3} involves electron states with
spin-up and spin-down projections. Electron-electron interaction
can flip electron spin. Therefore, a charge density wave state is
characterized by an order parameter
$\Delta_{p_1p_2}^{\sigma_{1}\sigma_{2}}(\textbf{Q})$ that is
matrix in the space of Landau level and spin indices. However, as
it will be clear from discussion below, if the Landau levels are
spin-resolved, i.e. $\Delta_{\rm ex} \gg \max\{T,\tau^{-1}\}$, the
charge density wave state creates only on the $N$th Landau level
with certain spin projection. Then Landau levels with different
spin projection become completely separated and can be ignored.
Thus, we can consider the charge density wave order parameter to
be matrix only in the space of Landau level indices. It is related
with distortion of electron density on the $(N-1)$th, $N$th, and
$(N+1)$th Landau levels as
\begin{equation}
\langle \rho(\textbf{q})\rangle =S
n_{L}\sum\limits_{p_1,p_2=N-1}^{N+1} \Delta_{p_1p_2}(\textbf{q})
F_{p_1p_2}(\textbf{q}), \label{rhoc3}
\end{equation}
where $S$ stands for the area of two-dimensional electron gas and
form-factor $F_{p_1p_2}(\textbf{q})$ is defined as
\begin{equation}
F_{p_1p_2}(\textbf{q})= n_L^{-1} \sum_k
\phi^{*}_{p_1k}(0)\phi_{p_2k}(\textbf{q}l_H^2) \exp\left
(\frac{i}{2} q_x q_y l_H^2\right ).  \label{FFc3}
\end{equation}

After Hartree-Fock decoupling of interaction term in effective
action (\ref{S3Lc3}) (see Ref.~\cite{FPA}), we obtain
\begin{gather}
\mathcal{S}=-\frac{N_{r}\Omega _{\Delta }}{T}+\int
d\textbf{r}\sum\limits_{p_1,p_2}\sum\limits_{\alpha,\omega_n,\omega_m}
\overline{\psi _{p_1\omega_{n}}^{\alpha}}(\textbf{r})
\Bigl[(i\omega_{n}+\mu\notag \\ - V_{\rm
dis}(\textbf{r}))\delta_{n m} -\mathcal{\hat H}  +
\lambda_{p_1p_2}(\textbf{r}) \Bigr]\psi
_{p_2\omega_{m}}^{\alpha}(\textbf{r}),\label{SHF1c3}
\end{gather}
where
\begin{equation}
\Omega _{\Delta }=\frac{n_{L}S^{2}}{2}\!\sum_{p_i}\!
\int\!\!\frac{d
\textbf{q}}{(2\pi)^2}U_{p_1p_2p_3p_4}(\textbf{q})\Delta_{p_1p_4}
(\textbf{q})\Delta_{p_3p_2}(-\textbf{q}). \label{SHF2c3}
\end{equation}
Potential $\lambda_{p_1p_2}(\textbf{r})$ in Eq.~\eqref{SHF1c3}
appears as a consequence of distortion of uniform electron density
by the charge density wave and is related with the order parameter
as
\begin{equation}
\lambda_{p_1p_2}(\textbf{q})=S \sum_{p_3p_4}
\frac{U_{p_3p_4p_1p_2}(\textbf{q})}{F_{p_1p_2}(-\textbf{q})}\Delta_{p_3p_4}
(\textbf{q}), \label{lambdac3}
\end{equation}
where $U_{p_1p_2p_3p_4}(\textbf{q})$ denotes the generalized
Hartree-Fock potential
\begin{gather}\label{UHFc3}
U_{p_1p_2p_3p_4}(\textbf{q})=-n_L \Bigl ( U_{\rm
scr}(\textbf{q})F_{p_1p_2}(\textbf{q})F_{p_3p_4}(-\textbf{q})\\-\int\frac{d
\textit{\textbf{p}}}{(2\pi)^2n_L}e^{-i
\textbf{q}\textit{\textbf{p}} l_{H}^{2}} U_{\rm
scr}(\textit{\textbf{p}})F_{p_1p_4}(\textit{\textbf{p}})F_{p_3p_2}(-\textit{\textbf{p}})\Bigr
).\notag
\end{gather}


\subsection{\label{ch3.1.4}Average over disorder}


After standard average over the random potential $V_{\rm
dis}(\textbf{r})$ (see Ref.~\cite{ELK}), effective action
\eqref{SHF1c3} becomes
\begin{gather}
\mathcal{S}=-\frac{N_{r}\Omega _{\Delta }}{T}+ \int d\textbf{r}
\psi ^{\dagger }(\textbf{r})\left( i\omega +\mu -\mathcal{\hat H}
+\check{\lambda} +iQ\right) \psi (\textbf{r})\notag
\\-\frac{1}{2g}\int d\textbf{r} \tr Q^{2}(\textbf{r}),
\label{S2c3}
\end{gather}
where we introduce new field $Q(\textbf{r})$, that is unitary
matrix in Matsubara and replica spaces. For convenience we use the
following notation
\begin{equation}\label{lc3}
\psi^\dagger \check\lambda \psi = \sum \limits_{p_1p_2} \sum
\limits_{\alpha,\omega_n}
\overline{\psi^{\alpha}_{p_1\omega_n}}(\textbf{r})\lambda_{p_1p_2}(\textbf{r})
\psi^{\alpha}_{p_2\omega_n}(\textbf{r}).
\end{equation}

Let us recall that action (\ref{S2c3}) at zero temperature, i.e.
for $\omega _{n}\to 0$, and in the absence of the induced
potential $\check\lambda (\textbf{r})$ and the time-dependent
vector-potential $\textbf{a}$ has the following saddle-point
solution
\begin{equation}
Q_{\rm sp}=V^{-1}P_{\rm sp}V,\qquad (P_{\rm sp})_{nm}^{\alpha
\beta }=P_{\rm sp}^{n}\delta _{nm}\delta ^{\alpha \beta },
\label{Pnmc3}
\end{equation}
where $V$ is arbitrary global unitary rotation and $P_{\rm
sp}^{n}$ obeys the equation
\begin{equation}
P_{sp}^{n}=i g G^{\omega_n}(\textbf{r},\textbf{r}),\qquad
G^{\omega_n}(\textbf{r},\textbf{r}^{\prime})=
\sum\limits_{p=N-1}^{N+1}
G_{p}^{\omega_n}(\textbf{r},\textbf{r}^{\prime}). \label{sp5c3}
\end{equation}
Here Green function
$G_{p}^{\omega_n}(\textbf{r},\textbf{r}^{\prime})$ is as follows
\begin{gather}\label{G0c3}
G_{p}^{n}(\textbf{r},\textbf{r}^{\prime})=\sum_{k}\phi _{pk}^{\ast
}(\textbf{r})G_{p}(\omega_{n})\phi_{pk}(\textbf{r}^{\prime}),\\
G_{p}(\omega_{n})=[i\omega_{n}+\mu_N +
\epsilon_N-\epsilon_{p}+iP_{\rm sp}^{n}]^{-1},\notag
\end{gather}
where chemical potential $\mu_N$ is measured from the $N$th Landau
level. The $\epsilon _{p}=\omega _{H}(p+1/2)$ and $\phi
_{pk}(\mathop{\bf r}
\nolimits)$ are the eigenvalues and eigenfunctions of the hamiltonian $%
\mathcal{H}$, and $k$ denotes pseudomomentum. In the case of weak
disorder, $\omega_H \tau \gg 1$, solution of Eq.(\ref{sp5c3})
yields
\begin{equation}
P_{\rm sp}^{n}=\frac{\sgn \omega_{n}}{2\tau}, \qquad
\frac{1}{2\tau}=\sqrt{g n_L}, \label{sp51c3}
\end{equation}
with $n_L = 1/2\pi l_H^2$.

The fluctuations of the $V$ field are responsible for the
localization corrections to the conductivity (in the weak
localization regime they correspond to the maximally crossed
diagrams). However, in the considered case, these corrections are
of the order of $\ln N /N\ll 1$ and, therefore, can be neglected.
For this reason we simply put $V=1$.

The presence of the induced potential $\check\lambda(\textbf{r})$
and the time-dependent vector potential $\textbf{a}$ results in a
shift of the
saddle-point value \eqref{sp51c3} due to the coupling to the fluctuations $%
\delta P=P-P_{\rm sp}$ of the $P$ field. The corresponding
effective action for the $\delta P$ field follows from Eq.
(\ref{SHF2c3}) after integrating out fermions:
\begin{gather}
\mathcal{S} =\int d\textbf{r}\tr \ln G^{-1}-\frac{N_{r}\Omega
_{\Delta }}{T}-\frac{1}{2g}\int d\textbf{r}\tr (P_{\rm sp}+\delta
P)^{2} \notag \\ + \int d\textbf{r}\tr\ln \Bigl[1+(i\delta P+\hat
K + \check\lambda )G\Bigr]. \label{SHF3c3}
\end{gather}
Finally, the thermodynamic potential can be written as
\begin{equation}
\Omega =-\frac{T}{N_{r}}\ln \int \mathcal{D}[\delta P]I[\delta
P]\exp \mathcal{S}, \label{Omega}
\end{equation}
where following Ref.\cite{P1} the integration measure $I[\delta
P]$ is given by
\begin{equation}
\ln I[\delta P]=-\frac{1}{(\pi \rho )^{2}}\int
\sum\limits_{nm}^{\alpha \beta }\left[ 1-\Theta (nm)\right] \delta
P_{nn}^{\alpha \alpha }\delta P_{mm}^{\beta \beta },  \label{mesI}
\end{equation}
with $\rho$ being the thermodynamic density of states and
$\Theta(x)$ the Heaviside step function.

The quadratic in $\delta P$ part of the action (\ref{SHF3c3})
together with the contribution (\ref{mesI}) from the integration
measure determine the propagator of the $\delta P$ fields (see
Ref.~\cite{Burm1} for details)
\begin{gather}
\langle \delta P_{m_{1}m_{2}}^{\alpha \beta }(\textbf{q})\delta
P_{m_{3}m_{4}}^{\gamma \delta }(-\textbf{q})\rangle \label{Pcor3c3} \\
=\frac{g \delta _{m_{1}m_{4}}\delta _{m_{2}m_{3}}\delta ^{\alpha
\delta }\delta ^{\beta \gamma }\Theta
(m_{1}m_{3})}{1+g\pi^{\omega_{m_{1}}}(\omega_{m_{3}}-\omega_{m_{1}},\textbf{q})}
-
\frac{2\left[ 1-\Theta (m_{1}m_{3})\right] }{(\pi \rho )^{2}}\notag \\
\times  \frac{g\delta _{m_{1}m_{2}}\delta ^{\alpha \beta
}}{1+g\pi^{\omega_{m_{1}}}(0,\textbf{q})}
  \frac{g\delta _{m_{3}m_{4}}\delta^{\delta \gamma
}}{1+g\pi^{\omega_{m_{3}}}(0,\textbf{q})},\notag
\end{gather}
where the bare polarization operator
$\pi^{\omega_m}(\nu_n,\textbf{q})$ involves Green functions for
the $(N-1)$th, $N$th, and $(N+1)$th Landau levels only
\begin{eqnarray}
\pi^{\omega_m}(\nu_n,\textbf{q})&=&\sum\limits_{p_1p_2}\pi_{p_1p_2}^{\omega_m}(\nu_n,\textbf{q})
=-n_{L}\sum\limits_{p_1p_2}G_{p_2}(\omega_{m})
\notag\\
&\times &G_{p_1}(\omega_{m}+\nu_{n})
F_{p_1p_2}(\textbf{q})F_{p_2p_1}(-\textbf{q}).\label{pi00c3}
\end{eqnarray}

%
%
\subsection{\label{ch3.1.5}Thermodynamic potential. Second order contribution}
%

In the absence of the time-dependent vector potential $\textbf{a}$
effective action \eqref{SHF3c3} should contain only
$\Delta_{NN}(\textbf{q})\equiv \Delta(\textbf{q})$ order parameter
in the limit $\max\{T,\tau^{-1}\}\ll \Delta_{\rm ex}\ll\omega_H$.
To demonstrate it, we find the second order contribution to the
thermodynamic potential for $\textbf{a}=0$.

Performing evaluation similar to one presented in
Ref.~\cite{Burm2}, we obtain
\begin{equation}
\Omega =\Omega^{(0)}+\Omega^{(2)}+\cdots, \label{Omega1c3}
\end{equation}
where
\begin{equation}
\Omega^{(0)}(\mu )=\int d\textbf{r}\tr\ln G^{-1}-\frac{1}{2g} \int
d\textbf{r}\tr P_{\rm sp}^{2} \label{Omega0c3}
\end{equation}
is the thermodynamic potential of the liquid state and
\begin{eqnarray}
\Omega^{(2)} &=&\frac{n_L S^2}{2 T} \sum\limits_{p_1\cdots p_4}
\int \frac{d \textbf{q} }{(2\pi)^2} \Biggl [
U_{p_1p_2p_3p_4}(\textbf{q})
 \label{dOmegac3}\\ &-& T
\sum\limits_{\omega_n}\sum\limits_{p_5\cdots p_8}
\frac{U_{p_1p_2p_5p_6}(\textbf{q})}{F_{p_5p_6}(\textbf{q})}
\frac{U_{p_3p_4p_7p_8}(-\textbf{q})}{F_{p_7p_8}(-\textbf{q})}\notag\\
&\times &\left (\delta_{p_5p_8}\delta_{p_6p_7} - \frac{g
\pi^{\omega_n}_{p_8p_7}(0,\textbf{q})}{1+g
\pi^{\omega_n}(0,\textbf{q})}\right
)\pi^{\omega_n}_{p_5p_6}(0,\textbf{q}) \Biggr ]\notag \\ &\times &
\Delta_{p_1p_2}(\textbf{q})\Delta_{p_3p_4}(-\textbf{q})\notag
\end{eqnarray}
is the contribution to the thermodynamic potential quadratic in
the order parameter $\Delta_{p_1p_2}(\textbf{q})$.

It is worthwhile to mention that polarization operators
$\pi^{\omega_n}_{p_1p_2}(\nu_n,\textbf{q})$ obey the following
hierarchy with respect to small parameter,
$\max\{T,\tau^{-1}\}/\omega_H \ll 1$,
\begin{equation}\label{pippc3}
\pi^{\omega_n}_{p_1p_2} \sim
  \begin{cases}
    \mathcal{O}(1), & p_1=p_2=N, \\
    \mathcal{O}\left (\frac{\max\{T,\tau^{-1}\}}{\omega_H} \right ), & p_1=N \,\text{or}\, p_2=N,\\
     \mathcal{O}\left (\left [\frac{\max\{T,\tau^{-1}\}}{\omega_H}\right ]^2 \right ),
     & p_1\neq N\,\text{and}\, p_2\neq N.
  \end{cases}
\end{equation}
According to the hierarchy \eqref{pippc3} we can write
\begin{equation}\label{approxpic3}
\pi_{p_1p_2}^{\omega_n}(0,\textbf{q}) \approx
\pi_{0}^{\omega_n}(0,q)\delta_{p_1N}\delta_{p_2N},
\end{equation}
where we introduce $\pi_{0}^{\omega_n}(0,q)\equiv
\pi_{NN}^{\omega_n}(0,q)$. Thus from Eq.\eqref{dOmegac3} we obtain
\begin{eqnarray}
\Omega^{(2)} &=&\frac{n_L S^2}{2 T} \sum\limits_{p_1\cdots p_4}
\int \frac{d \textbf{q} }{(2\pi)^2} \Biggl [
U_{p_1p_2p_3p_4}(\textbf{q}) \label{dOmega1c3}\\
 &+&T \sum\limits_{\omega_n} \frac{n_L G^2_N(\omega_n)}{1+g
\pi^{\omega_n}_0(0,\textbf{q})} U_{p_1p_2NN}(\textbf{q})
U_{p_3p_4NN}(-\textbf{q})\Biggr ]\notag \\
&\times
&\Delta_{p_1p_2}(\textbf{q})\Delta_{p_3p_4}(-\textbf{q}).\notag
\end{eqnarray}

To find the possible non-zero order parameters
$\Delta_{p_1p_2}(\textbf{q})$, we should diagonalize the $9\times
9$ matrix in Eq.\eqref{dOmega1c3}. Fortunately, non-trivial part
of Eq.\eqref{dOmega1c3} can be written as
\begin{eqnarray}
\delta \Omega^{(2)} &=& \frac{n_L S^2}{2 T} \int \frac{d
\textbf{q} }{(2\pi)^2} T_0(q) \left (\Delta(\textbf{q}),\,\,
\frac{T_1(q)}{T_0(q)}\varphi(\textbf{q})\right )\notag \\ &\times
&
\begin{pmatrix}
  a(q) & a(q)  \\
  a(q) & a(q)+2 \xi(q)
\end{pmatrix}  \begin{pmatrix}
 \Delta(-\textbf{q}) \\
\displaystyle \frac{T_1(q)}{T_0(q)}\varphi(-\textbf{q})
\end{pmatrix}.\label{dOmega2c3}
\end{eqnarray}
Here $\varphi(\textbf{q})$ involves a linear combination of all
order parameters $\Delta_{p_1p_2}(\textbf{q})$ except
$\Delta(\textbf{q})$. Characteristic energies $T_0(q)$ and
$T_1(q)$ is related with Hartree-Fock potential \eqref{UHFc3} as
follows
\begin{equation}\label{T0T1c3}
T_0(q) = \frac{U_{NNNN}(q)}{4},\qquad T_1(q) = e^{i\phi}
\frac{U_{N,N\pm 1,NN}(\textbf{q})}{4},
\end{equation}
where $\phi$ denotes angle of vector $\textbf{q}$ with respect to
the $x$ axis. We emphasize that quantity $T_1(q)$ depends only on
the absolute value $q$ of vector $\textbf{q}$. Matrix element
$a(q)$ is given by
\begin{equation}\label{aTqc3}
a(T,\tau^{-1},q) = 1 + 4 T \sum\limits_{\omega_n} \frac{n_L T_0(q)
G^2_N(\omega_n)}{1+g \pi^{\omega_n}(0,q)},
\end{equation}
whereas function $\xi(q)$ is defined as
\begin{equation}\label{xiqc3}
\xi(q) = \frac{T_0(q)}{2 T_1(q)}-\frac{1}{2}.
\end{equation}

The eigenvalues of the $2\times 2$ matrix in Eq.\eqref{dOmega2c3}
can be easily found
\begin{equation}\label{eigenc3}
\lambda_{\pm}(q) = a(q) +\xi(q) \pm \sqrt{[a(q)]^2 +[\xi(q)]^2}
\end{equation}
As one can check, the eigenvalue $\lambda_{+}(q)$ has the same
sign as $\xi(q)$ for all values of $a(q)$ whereas the eigenvalue
$\lambda_{-}(q)$ changes its sign at point $a(q)=0$. Therefore,
the instability appears at the same condition as if we consider
only one charge density wave order parameter $\Delta(\textbf{q})$
as it has usually done ~\cite{KFS,FKS,MC,Burm2}. According to the
result derived in Appendix~\ref{ch3.App.1}, characteristic energy
$T_1(q)$ is of the order of $T_0(q)/N \ll T_0(q)$. By using the
condition $\xi(q) \gg a(q)$, we find therefore
\begin{eqnarray}
\Omega^{(2)} &=& \frac{n_L S^2}{2 T} \int \frac{d \textbf{q}
}{(2\pi)^2} T_0(q) \Biggl [ a(q)\left (1-\frac{T_1(q)}{T_0(q)}
a(q)\right )\notag\\ &\times &
\Delta_{-}(\textbf{q})\Delta_{-}(-\textbf{q}) +
\frac{T_1(q)}{T_0(q)}
\Delta_{+}(\textbf{q})\Delta_{+}(-\textbf{q}) \Biggr
],\label{dOmega3c3}
\end{eqnarray}
where
\begin{eqnarray}\label{dpm3}
\Delta_{-}(\textbf{q}) &=& \Delta(\textbf{q}) - \left
(\frac{T_1(q)}{T_0(q)}\right )^2 a(q) \varphi(\textbf{q}), \\
\Delta_{+}(\textbf{q}) &=& \varphi(\textbf{q}) +
a(q)\Delta(\textbf{q}).\notag
\end{eqnarray}
Minimum of the free energy~\cite{Foot2} is reached at
$\Delta_{+}(q)=0$. Neglecting the difference of the order of
$\mathcal{O}(N^{-2})$ between $\Delta_{-}(q)$ and $\Delta(q)$, we
obtain finally
\begin{eqnarray}
\Omega^{(2)} &=& \frac{n_L S^2}{2 T} \int \frac{d \textbf{q}
}{(2\pi)^2} T_0(q) a(q)\left (1-\frac{T_1(q)}{T_0(q)} a(q)\right
)\notag\\ &\times &
\Delta(\textbf{q})\Delta(-\textbf{q}).\label{dOmega4c3}
\end{eqnarray}
Thus, the fact that the order parameters
$\Delta_{p_1p_2}(\textbf{q})$ with $p_1$ and $p_2$ different from
$N$ can exist leads to correction of the order of
$\mathcal{O}(N^{-1})$. Later on we assume therefore that
\begin{equation}\label{ordpc3}
\Delta_{p_1p_2}(\textbf{q}) =
\Delta(\textbf{q})\delta_{p_1N}\delta_{p_2N}.
\end{equation}

%
%
\subsection{\label{ch3.1.6}``Three-level'' model}
%

The results of the previous section allows us to establish finally
an effective action for the ``three-level'' model.

According to definition \eqref{lambdac3}, charge density wave on
the $N$th Landau level with the order parameter
$\Delta(\textbf{q})$ results in the induced potential
$\lambda_{N,N\pm 1}(\textbf{q})$, scattering electrons from the
$N$th Landau level to the $(N\pm 1)$th Landau level. However, the
induced potential is of the order of $T_1(q)$ and, consequently,
leads to small corrections of the order of $\mathcal{O}(N^{-1})$.
For the reasons to be explained shortly, we write
\begin{equation}
\lambda_{p_1p_2} (\textbf{q})=S U(q)F_{N}^{-1}(q)\Delta
(\textbf{q})\delta_{p_1N}\delta_{p_2N}. \label{lambda23c3}
\end{equation}

Finally, the effective action for the ``three-level'' model
becomes
\begin{eqnarray}
\mathcal{S}_{TL}[\delta P] &=&\int d\textbf{r}\tr \ln
G^{-1}-\frac{N_{r}\Omega _{\Delta }}{T}\notag \\
&-& \frac{1}{2g}\int d\textbf{r}\tr (P_{\rm sp}+\delta P)^{2}
\label{SHF5c3}\\ &+& \int d\textbf{r}\tr\ln \Bigl[1+(i\delta
P+\hat K + P_N\lambda P_N )G\Bigr],\notag
\end{eqnarray}
where
\begin{eqnarray}
P_{N}(\textbf{r}_{1},\textbf{r}_{2}) &=& \sum_{k} \phi^{*}_{N
k}(\textbf{r}_{2}) \phi_{N k}(\textbf{r}_{1}) \notag \\ &=& n_L
\exp \left (i
\frac{(y_1-y_2)(x_1+x_2)}{2l_H^2}\right )\label{pn} \\
&\times &\exp \left (
-\frac{|\textbf{r}_1-\textbf{r}_2|^2}{4l_H^2}\right ) L_N\left
(\frac{|\textbf{r}_1-\textbf{r}_2|^2}{2l_H^2}\right )\notag
\end{eqnarray}
is the projection operator on the $N$th Landau level ($L_N(x)$
denotes the Laguerre polynomial) and
\begin{equation}
\Omega _{\Delta }=\frac{n_{L}S^{2}}{2}\int\frac{d
\textbf{q}}{(2\pi)^2}U(q)\Delta(\textbf{q})\Delta(-\textbf{q}),
\label{SHF6c3}
\end{equation}
Here, for a brevity we introduce $U(q)\equiv U_{NNNN}(q)$.

%
%
\section{\label{ch3.2} Conductivity of the UCDW state at $T_c-T\ll T_c$}
%

%
\subsection{\label{ch3.2.1} Conductivity tensor $\sigma_{ab}$}

Effective action \eqref{SHF5c3} allows us to evaluate conductivity
of the system in the CDW state. As the most interesting case we
consider the half-filled $N$th Landau level where at $T<T_c$ the
UCDW state exists. The order parameter is~\cite{FPA}
\begin{equation}
\Delta (\textbf{q})=\frac{(2\pi )^{2}}{S}\Delta \Bigl[\delta
(\textbf{q}-\textbf{Q}_0)+\delta (\textbf{q}-\textbf{Q}_0)\Bigr],
\label{DeltaU}
\end{equation}
where vector $\textbf{Q}_0$ that determines period and direction
of the UCDW state can be oriented along spontaneously chosen
direction. Usually, its direction is fixed either by intrinsic
anisotropy of the system or by small magnetic field applied
parallel to 2DEG~\cite{PDSTPBW,LCEPW2,CLEJPW,CEPW,ZPSPW}. We
assume that the vector $\textbf{Q}_0$ is directed at an angle
$\phi$ with respect to the $x$ axis. Let us recall that the
absolute value of the vector $\textbf{Q}_0$ equals
$Q_{0}=r_{0}/R_{c}$ with $r_{0}\approx 2.4$ being the first zero
of the Bessel function $\mathcal{J}_0(x)$.

The conductivity tensor $\sigma_{ab}(\nu_n,q)$ at $q=0$ can be
found after integration over $\delta P(\textbf{r})$ fields as the
second derivative of logarithm of the effective action with
respect to spatially constant time-dependent vector potential
$\textbf{a}(\nu_n)$,
\begin{eqnarray}\label{sigmadefinitionc3}
\sigma_{ab}(\nu_n) &=& \frac{\pi T}{S N_r \nu_n} \frac{\delta^2
}{\delta a_a(\nu_n)\delta a_b(-\nu_n)}\\ &\times & \ln \int
\mathcal{D}\delta P\, I[\delta P] \exp \mathcal{S}_{TL}[\delta P]
\Biggl |_{\textbf{a} =0}.\notag
\end{eqnarray}

It is worthwhile to mention that Eq.\eqref{sigmadefinitionc3}
corresponds to the term $\textbf{j}(\nu_n) \textbf{a}(-\nu_n)$
with $\textbf{j}(\nu_n)$ being current density in the effective
action for the vector potential $\textbf{a}(\nu_n)$. As one can
check by inspection, contribution to the conductivity tensor of
the first order in the order parameter $\Delta(\textbf{q})$
vanishes. It occurs because the UCDW state appears at non-zero
vector $\textbf{Q}_0$. Thus, the first non-vanishing contribution
to the conductivity tensor of the UCDW state is of the second
order in the order parameter $\Delta$.

\begin{figure}[t]
\centerline{\includegraphics[width=70mm]{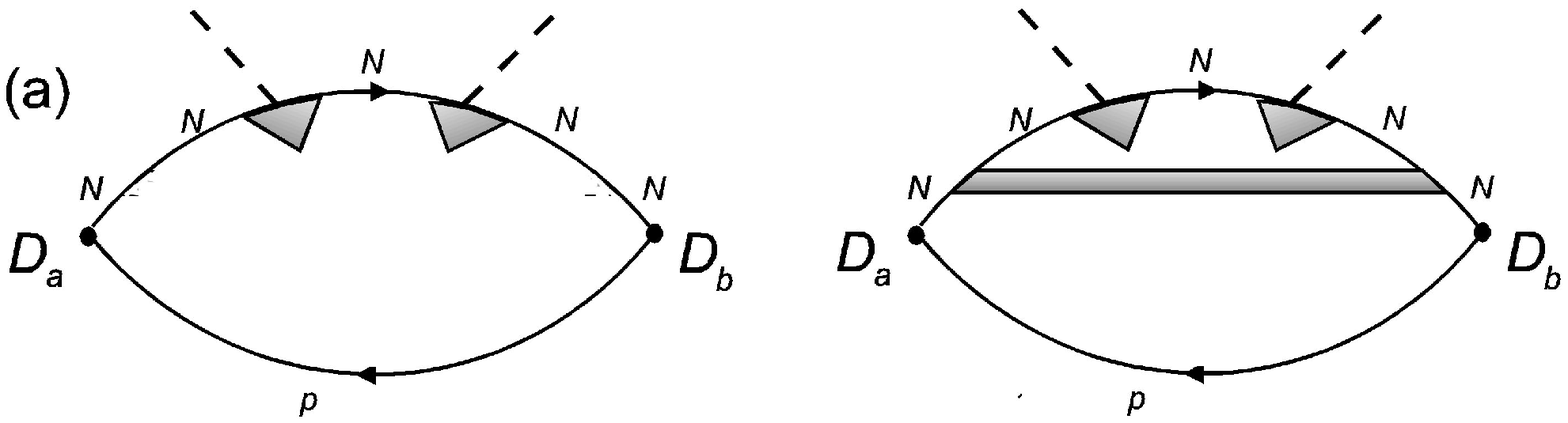}}
\centerline{\includegraphics[width=70mm]{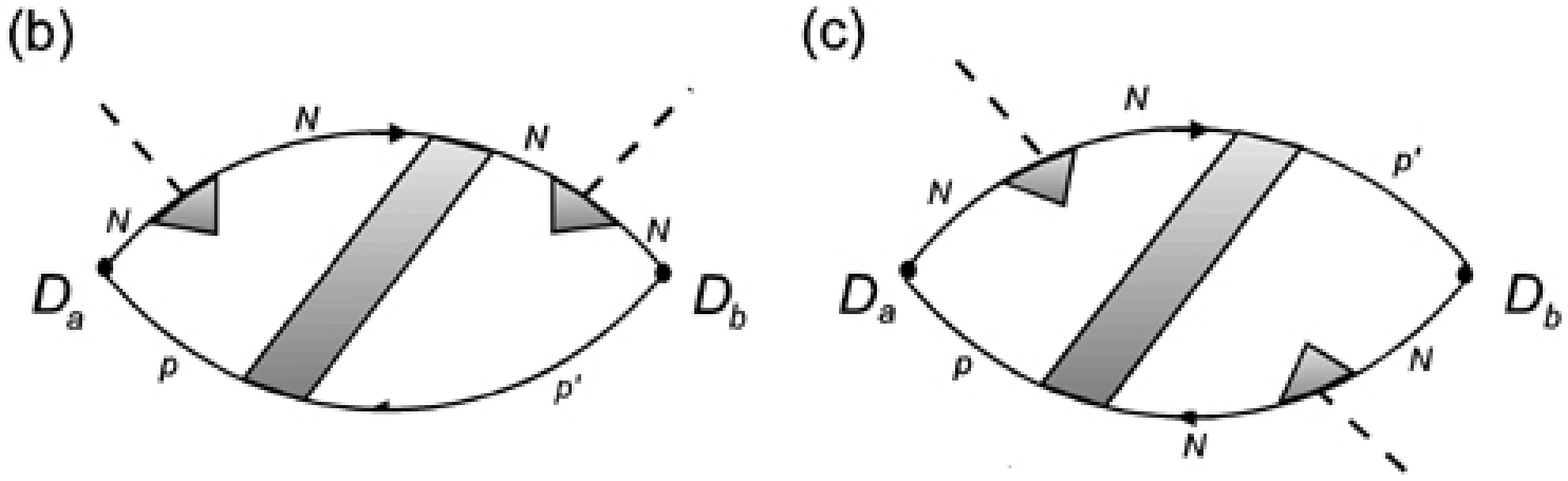}}
\caption{Diagrams for corrections to the conductivity tensor
$\sigma_{ab}$. Solid line denote Green function, $N / p /
p^{\prime}$ Landau level indices, dashed line the induced
potential $\lambda(\textit{\textbf{\textit r}})$, and shaded block
the impurity ladder.} \label{FIG4.1}
\end{figure}

In order to find it, we expand the effective action
$\mathcal{S}_{TL}[\delta P]$ to the second order in both the
induced potential $\lambda(\textbf{r})$ and the $\hat K$. Then, we
integrate over $\delta P(\textbf{r})$ fields. We do not present
the explicit calculations here since they are similar to ones
presented in the Ref.~\cite{Burm2}. We mention that there are
three contributions of different structure to the conductivity
tensor of the UCDW state. Diagrams for them are shown in
Fig.~\ref{FIG4.1}.
\begin{figure}[t]
\centerline{\includegraphics[width=70mm]{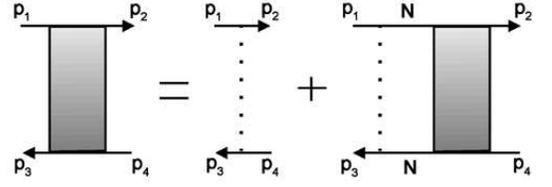}}
\caption{Impurity ladder. Frequency $\omega_n+\nu_n$ runs from
right to left, whereas $\omega_n$ runs from left to right.}
\label{FIG4.2}
\end{figure}

The first and the second diagrams (Fig.~\ref{FIG4.1}(a))
correspond to the following contribution
\begin{eqnarray}
\sigma_{ab}^{(a)}(\nu_n)&=& -  \frac{8\pi\omega_H}{m \nu_n} T_0^2
\Delta^2 T \sum\limits_{\omega_n} \frac{G_N^3(\omega_n)}{1+g
\pi^{\omega_n}_0(0,0)}\notag \\
&\times &\sum \limits_p \frac{D_{Np}^a D_{pN}^b
G_p(\omega_n+\nu_n)}{(1+g \pi^{\omega_n}_0(0,Q_0))^2}.
\label{sigmaAc3}
\end{eqnarray}
Here $D_{Np}^{a}$ denotes matrix element of the covariant
derivative
\begin{eqnarray}
D_{Np}^{a} &=& \int d \textbf{r} \phi^{*}_{Nk}(\textbf{r}) D_a
\phi_{pk}(\textbf{r})=\sqrt{n_L}\Bigl [ \delta_{p,N-1} \beta^{a} \sqrt{N} \notag \\
 &+& \delta_{p,N+1} \gamma^{a}
\sqrt{N+1}\Bigr ],\label{matelemc3}
\end{eqnarray}
where
\begin{equation}\label{gbc3}
\gamma^{x} = i, \quad \gamma^{y} = 1,\quad \beta^{x}=-i,\quad
\beta^{y}=1.
\end{equation}
We note that Eq.\eqref{sigmaAc3} contains only isotropic
contribution to the conductivity tensor, i.e.
$\sigma_{xx}^{(a)}=\sigma_{yy}^{(a)}$ and
$\sigma_{xy}^{(a)}=-\sigma_{yx}^{(a)}$ for some direction of the
UCDW.


\begin{table*}
\begin{ruledtabular}
\centerline{\begin{tabular}{||rcl||rcl||}
  $I_{N,N-1,N-1,N}$& $=$ & $g n_L \mathcal{J}_0^2(r_0)$ &
  $I_{N,N+1,N+1,N}$ & $=$ & $g n_L \mathcal{J}_0^2(r_0)$ \\
  $I_{N,N-1,N,N+1}$ & $=$ & $g n_L \mathcal{J}_1^2(r_0)$ &
  $I_{N,N+1,N,N-1}$ & $=$ & $g n_L \mathcal{J}_1^2(r_0)$ \\
  $I_{N,N,N-1,N}$ & $=$ & $g n_L e^{i\phi}\mathcal{J}_0(r_0)\mathcal{J}_1(r_0)$ &
  $I_{N,N+1,N,N}$ & $=$ & $g n_L e^{-i\phi}\mathcal{J}_0(r_0)\mathcal{J}_1(r_0)$ \\
  $I_{N,N-1,N,N}$ & $=$ & $-g n_L e^{i\phi}\mathcal{J}_0(r_0)\mathcal{J}_1(r_0)$ &
  $I_{N,N,N+1,N}$ & $=$ & $-g n_L e^{-i\phi}\mathcal{J}_0(r_0)\mathcal{J}_1(r_0)$ \\
  $I_{N,N-1,N,N-1}$ & $=$ & $g n_L e^{2 i\phi}\mathcal{J}_1^2(r_0)$ &
  $I_{N,N+1,N,N+1}$ & $=$ & $g n_L e^{-2 i\phi}\mathcal{J}_1^2(r_0)$ \\
  $I_{N,N-1,N+1,N}$ & $=$ & $g n_L e^{2 i\phi}\mathcal{J}_1(r_0)\mathcal{J}_0(r_0)$ &
    $I_{N,N+1,N-1,N}$ & $=$ & $g n_L e^{-2 i\phi}\mathcal{J}_1(r_0)\mathcal{J}_0(r_0)$ \\
    \end{tabular}}
\vspace{1cm} \caption{Expressions for quantities
$I_{p_1p_2p_3p_4}(\textbf{Q}_0)$ involved in
Eqs.\eqref{sigmaAc3},\eqref{sigmaBc3} and \eqref{sigmaCc3}.}
\label{Tablec31}
\end{ruledtabular}
\end{table*}
The third diagram (Fig.~\ref{FIG4.1}(b)) is given by
\begin{eqnarray}
\sigma_{ab}^{(b)}(\nu_n)&=& \frac{8\pi\omega_H}{\nu_n m}
  T_0^2\Delta^2 T \sum\limits_{\omega_n}\sum \limits_{p
p^{\prime}} \frac{G^4_N(\omega_n+\nu_n)}{(1+g
\pi^{\omega_n}_0(0,Q_0))^2}\notag\\&\times &D_{pN}^a
D_{Np^{\prime}}^b G_p(\omega_n)G_{p^{\prime}}(\omega_n) \Bigl [
I_{Npp^{\prime}N}(\textbf{Q}_0) \notag \\ &+& \frac{n_L
G_N(\omega_n)G_N(\omega_n+\nu_n)}{1+g
\pi^{\omega_n}_0(\nu_n,Q_0)}\notag \\ &\times &
I_{NpNN}(\textbf{Q}_0)I_{NNp^{\prime}N}(\textbf{Q}_0)\Bigr ].
\label{sigmaBc3}
\end{eqnarray}
Symbol $I_{p_1p_2p_3p_4}(\textbf{Q})$ denotes the impurity ladder
in the Landau level index representation (see Fig.~\ref{FIG4.2})
\begin{equation}\label{Ic3}
I_{p_1p_2p_3p_4}(\textbf{Q}) = g \int_q F_{p_1p_2}(\textbf{q})
F_{p_3p_4}(-\textbf{q}) \exp\left (-i \textbf{q} \textbf{Q}
l_H^2\right ).
\end{equation}
Evaluation of $I_{p_1p_2p_3p_4}(\textbf{Q}_0)$ is given in
Appendix~\ref{ch3.App.2}. For convenience we present the results
for quantities $I_{p_1p_2p_3p_4}(\textbf{Q}_0)$ in the
Table~\ref{Tablec31}. As one can see, the
$I_{NNNpN}(\textbf{Q}_0)$, $I_{NpNNN}(\textbf{Q}_0)$ and
$I_{Npp^{'}N}(\textbf{Q}_0)$ are proportional to
$\mathcal{J}_0(r_0)=0$. Thus, the contribution \eqref{sigmaBc3}
vanishes
\begin{equation}\label{sB1c3}
\sigma_{ab}^{(b)}(\nu_n) = 0.
\end{equation}

The last diagram (Fig.~\ref{FIG4.1}(c)) can be written as
\begin{eqnarray}
\sigma_{ab}^{(c)}(\nu_n)&=& \frac{8\pi\omega_H}{\nu_n m}
T_0^2\Delta^2
 T \sum\limits_{\omega_n}\sum \limits_{p
p^{\prime}} \frac{G_N^2(\omega_n)G^2_N(\omega_n+\nu_n)}{(1+g
\pi^{\omega_n}_0(0,Q_0))}\notag\\&\times &
\frac{G_p(\omega_n)G_{p^{\prime}}(\omega_n+\nu_n)}{(1+g
\pi^{\omega_n+\nu_n}_0(0,Q_0))} \notag \\ &\times &\frac{D_{pN}^a
D_{p^{\prime}N}^bI_{NpNp^{\prime}}(\textbf{Q}_0)}{1+g
\pi^{\omega_n}_0(\nu_n,Q_0)}. \label{sigmaCc3}
\end{eqnarray}

We note that terms in the sum over Landau level indices with
$p=p^{'}=N\pm 1$ lead to the anisotropic contribution. Terms with
$p=N\pm 1$ and $p^{'}=N\mp 1$ result in the isotropic
contribution.

%
\subsection{\label{ch3.2.2}Anisotropic contribution $\sigma_{ab}^{(\rm anis)}$}

\begin{figure}[t]
\centerline{\includegraphics[width=70mm]{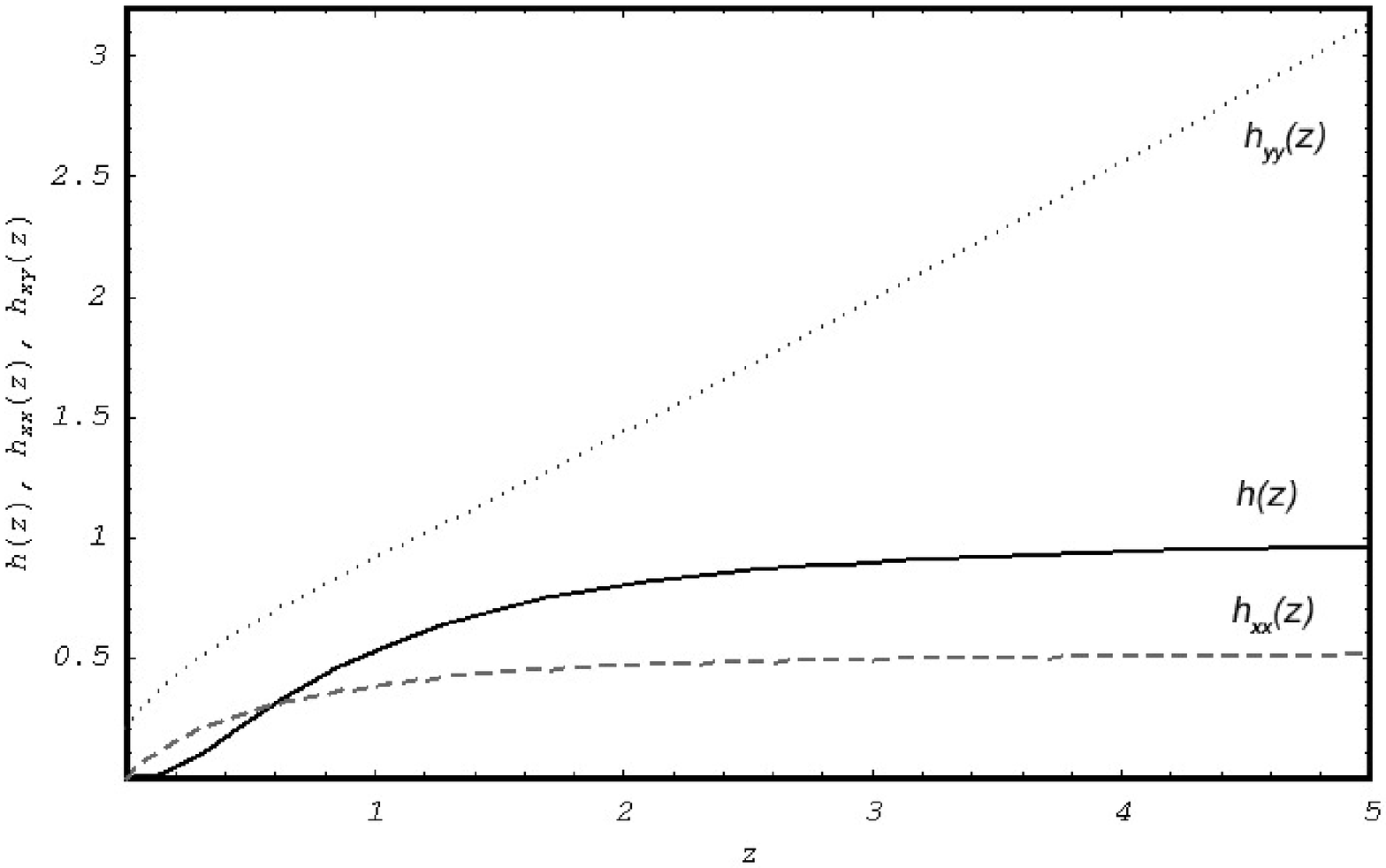}}
\caption{Functions $h(z)$, $h_{xx}(z)$ and $h_{xy}(z)$.}
\label{FIG4.3}
\end{figure}

We start our analysis of general expressions obtained in the
previous section from Eq.\eqref{sigmaCc3} that contains the
anisotropic contribution to the conductivity tensor. Taking into
account only terms with $p=p^{'}=N\pm 1$ in the sum over Landau
level indices involved in Eq.\eqref{sigmaCc3} we obtain
\begin{equation}\label{DCanis1c3}
\left . \begin{array}{c}
 \sigma_{xx}^{(\rm anis)} \\
 \sigma_{yy}^{(\rm anis)}
\end{array}
\right \} = \mp 4 \pi N  \mathcal{J}^2_1(r_0) h\left(\frac{1}{4\pi
T_c\tau}\right )\Delta^2 \cos[2\phi] ,
\end{equation}
and
\begin{equation}\label{DCanis2c3}
\left . \begin{array}{c}
 \sigma_{xy}^{(\rm anis)} \\
 \sigma_{yx}^{(\rm anis)}
\end{array}
\right \} = 4 \pi N \mathcal{J}^2_1(r_0) h\left(\frac{1}{4\pi
T_c\tau}\right )\Delta^2 \sin[2\phi].
\end{equation}
Function $h(z)$ is given as
\begin{equation}\label{hanisc3}
h(z) = 5 z^3 \frac{\zeta\left(6,\frac{1}{2}+z\right
)}{\left[\zeta\left(2,\frac{1}{2}+z\right )\right ]^2}=
  \begin{cases}
    \displaystyle \frac{4 \pi^2}{3} z^3, &\qquad  z \ll 1,\\
    \displaystyle 1 - \frac{3}{z}, &\qquad z \gg 1,
  \end{cases}
\end{equation}
where $\zeta(k,z)$ denotes the generalized Riemann
zeta-function~\cite{Foot3}. Function $h(z)$ increase monotonically
from $0$ to $1$, as it is shown in Fig.~\ref{FIG4.3}.

Eqs.\eqref{DCanis1c3} and \eqref{DCanis2c3} constitute one of the
main results of the present paper. Anisotropic contributions
\eqref{DCanis1c3} and \eqref{DCanis2c3} is seemed to be
proportional to $(T_c-T)/T_c$ since in the Landau theory the order
parameter $\Delta \propto \sqrt{(T_c-T)/T_c}$ (see
Ref.~\cite{Burm2} for explicit expression).

The $\sigma_{xx}^{(\rm anis)}$ as the function of angle $\phi$ has
the minimum at $\phi=0$ that corresponds to the vector
$\textbf{Q}_0$ directed along the $x$ axis. We note that the
modulation of electron density along the $y$ axis is absent in
this case. At $\phi=0$ conductivity $\sigma_{yy}^{(\rm anis)}$ is
positive whereas $\sigma_{xx}^{(\rm anis)}$ is negative, moreover,
they have the same absolute values. It leads to the statement that
the conductivity $\sigma_{xx}$ (along the electron density
modulation) is less than conductivity $\sigma_{yy}$ (across the
modulation). We emphasize that at $\phi=0$ conductivity
$\sigma_{xy}^{(\rm anis)}$ vanishes.

If the vector $\textbf{Q}_0$ is oriented at angle $\phi=\pi/4$
with respect to the $x$ axis conductivities $\sigma_{xx}^{(\rm
anis)}$ and $\sigma_{yy}^{(\rm anis)}$ vanish due to the symmetry
between the $x$ and $y$ axes. Vice versa, conductivity
$\sigma_{xy}^{(\rm anis)}$ reaches the maximum. It is worthwhile
to mention that anisotropic contributions to the conductivity of
UCDW state are proportional to $N$ as the conductivity of the
liquid state~\cite{AFS}.

%
\subsection{\label{ch3.2.3} Isotropic contribution $\sigma_{ab}^{(\rm isot)}$}

Eqs.\eqref{sigmaAc3} and \eqref{sigmaCc3} allows us to find also
the isotropic contribution to the conductivity tensor at $T_c-T\ll
T_c$. Taking into account Eq.\eqref{sigmaAc3} and terms with
$p=N\pm 1$ and $p^{'}=N\mp 1$ in the sum over Landau level indices
in Eq.\eqref{sigmaCc3}, we obtain the following isotropic
contribution to conductivity $\sigma_{xx}$
\begin{equation}\label{DCis1c3}
 \delta\sigma_{xx}^{(\rm isot)} = - 4 \pi N  h_{xx}\left (\frac{1}{4\pi T_c\tau}\right ) \Delta^2,
\end{equation}
where function $h_{xx}(z)$ is given by
\begin{eqnarray}\label{hxxc3}
h_{xx}(z) &=& \mathcal{J}^2_1(r_0) h\left (z\right )+\frac{1}{2 z
\zeta\left(2,\frac{1}{2}+z\right )}\\ &\times &\left [ 1 -
\frac{\Re \psi^{'}\left (\frac{1}{2}+(1-i)z\right
)}{\zeta\left(2,\frac{1}{2}+z\right )}\right ].\notag
\end{eqnarray}
Here $\psi(z)$ stands for the Euler di-gamma function. The
$h_{xx}(z)$ has asymptotic expressions as
\begin{equation}
h_{xx}(z)=  \begin{cases}
    \displaystyle\frac{2}{3} z, &\qquad z \ll 1, \\
    \displaystyle \mathcal{J}^2_1(r_0)+\frac{1}{4} - \frac{3 \mathcal{J}^2_1(r_0)-\frac{1}{4}}{z},&\qquad z \gg 1.
  \end{cases}
\end{equation}
As one can see from Eq.\eqref{DCis1c3}, isotropic contribution
$\delta\sigma_{xx}^{(\rm isot)}$ to conductivity $\sigma_{xx}$ is
of the same order as the anisotropic contribution.

The isotropic contribution to conductivity $\sigma_{xy}$ is as
follows
\begin{equation}\label{DCis2c3}
 \delta\sigma_{xy}^{(\rm isot)} = -8 \pi^2 N \frac{T_c}{\omega_H}
 h_{xy}\left (\frac{1}{4\pi T_c\tau}\right )\Delta^2,
\end{equation}
where
\begin{eqnarray}\notag
h_{xy}(z) &=&  \Biggl [\mathcal{J}^2_1(r_0) \left [ 4 z^4
\zeta\left (5,\frac{1}{2}+z\right )+z^3 \zeta\left
(4,\frac{1}{2}+z\right )\right ] \\ &-& 2 h_{xx}(z) + z \Im
\psi^{'}\left (\frac{1}{2}+(1-i)z\right ) \notag \\ &-& \Im
\psi\left (\frac{1}{2}+(1-i)z\right )\Biggr ] \notag \\ &\times &
\frac{1}{z \left [\zeta\left(2,\frac{1}{2}+z\right )\right ]^2} .
\label{hxyc3}
 \end{eqnarray}
The function $h_{xy}(z)$ has the following asymptotic expressions
in the limits of small and large $z$
\begin{equation}\label{hxyasc3}
 h_{xy}(z) =
  \begin{cases}
    \displaystyle\frac{2}{\pi^2}\left (1 - \frac{\pi^2 -2 \psi^{''}(\frac{1}{2})}{\pi^2} z\right ), &\qquad z \ll 1, \\
    \displaystyle \left (\frac{\pi}{4}-\frac{2\mathcal{J}^2_1(r_0)}{3}\right ) z, & \qquad
    z\gg 1.
    \end{cases}
\end{equation}
We mention that the isotropic contribution $\sigma_{xy}^{(\rm
isot)}$ contains additional small factor
$\max\{T_c,\tau^{-1}\}/\omega_H$ compared to the others. Results
\eqref{DCis1c3} and \eqref{DCis2c3} are one of the main results of
the paper.

%
\section{\label{ch3.3}Effect of the order parameter fluctuations on the conductivity tensor}

\subsection{\label{ch3.3.0}Order parameter fluctuations}

The order parameter $\Delta(\textbf{r})$ has meaning of the
saddle-point solution for a plasmon field that appears in the
Hubbard-Stratonovich transformation~\cite{HS} of the screened
electron-electron interaction. The expansion of such physical
quantities as free energy and linear response in the order
parameter series can be justified if fluctuations of the order
parameter can be neglected. As it was shown~\cite{Burm2},
fluctuations of the order parameter results in the first order
transition from the liquid state to the UCDW state at temperature
$T_c -\delta T_c$ where $\delta T_c/T_c \propto N^{-2/3} \ll 1$.
In the present section we investigate the effect of the
fluctuations on the conductivity tensor above and below the
mean-field transition.

In the previous section we assumed that the direction of the CDW
vector $\textbf{Q}_0$ is fixed by intrinsic anisotropy of crystal
or by applied parallel to 2DEG small magnetic field. However, the
functional dependence of anisotropy term in the hamiltonian was
insignificant for mean-field results obtained above. Now it should
be concretized. Experimental research of the anisotropy that
determines the direction along which the UCDW creates has been
performed in a number of
papers~\cite{PDSTPBW,LCEPW2,CLEJPW,CEPW,ZPSPW}. The results
obtained can be explained if one suggests that the Hartree-Fock
potential $U(\textbf{Q})$ involves terms proportional to $\cos
2\phi$ and $\cos 4\phi$. We mention that the term $\cos 2\phi$ can
be derived when small magnetic field parallel to 2DEG
applied~\cite{JMSG}. However, without parallel magnetic field the
term $\cos 2\phi$ is restricted by the symmetry of bulk $GaAs$
crystal. To date its physical origin is unknown~\cite{CLEJPW}. As
experimentally proven~\cite{ZPSPW}, coefficient of the $\cos
2\phi$ term depends on the density $n$ of electrons. Moreover, at
some certain value $n_{*}$ of the electron density it vanishes and
next term proportional to $\cos 4\phi$ becomes important. Below we
restrict our discussion to the general case $n\neq n_{*}$. We note
that the typical value of the anisotropy energy $E_A$ is of the
order of $1\, mK$ per electron as it is obtained from
experiment~\cite{CLEJPW}. In order to take into account the
anisotropy quantitatively we perform the following substitution
(see Eq.\eqref{T0T1c3})
\begin{equation}\label{flT0c3}
 T_0(Q)\to T_0(Q) + E_A\frac{1-\cos 2\phi}{2}
\end{equation}
near $Q=Q_0$. We note that the expression above has minimum at
$\phi=0$.

At $T> T_c$ the UCDW order parameter is zero in average $\langle
\Delta\rangle=0$ but the average of its square is non-zero
$\langle \Delta^2\rangle \neq 0$. It results in additional
contribution to the conductivity tensor of the liquid state. It is
worthwhile to mention that the contribution discussed above is
analogous to one for normal metal due to superconducting paring
above critical temperature~\cite{AL}.

The additional contribution to the conductivity tensor due to the
order parameter fluctuations can be found with a help of the
substitution $\langle
\Delta(\textbf{Q})\Delta(-\textbf{Q})\rangle$  for $\Delta^2$ in
Eqs.\eqref{DCanis1c3}, \eqref{DCanis2c3}, \eqref{DCis1c3} and
\eqref{DCis2c3} and averaging over all possible vectors
$\textbf{Q}$. The Green function of the order parameter is as
follows (see Ref.~\cite{Burm2})
\begin{eqnarray}
\langle\Delta(\textbf{Q})\Delta(-\textbf{Q})\rangle &=& \frac{T_c
}{4 T_0(Q_0) n_L} \Bigl [\frac{T-T_c}{T_c} + \gamma\left
(\frac{1}{4\pi T_c \tau}\right )\notag \\ &\times &(Q-Q_0)^2 R_c^2
+ \eta \sin^2\phi\Bigr ]^{-1},\label{fl1c3}
\end{eqnarray}
where dimensionless parameter $\eta=E_A /T_0(Q_0)$ and we
introduce
\begin{equation}\label{fl11c3}
\gamma(z) = \beta_1 + \mathcal{J}_1^2(r_0) z^2
\frac{\zeta(4,\frac{1}{2}+z)}{\zeta(2,\frac{1}{2}+z)}.
\end{equation}
Here constant $\beta_1 \approx 2.58$. After integration over
absolute value of vector $\textbf{Q}$ we find that in
Eqs.\eqref{DCanis1c3}, \eqref{DCanis2c3}, \eqref{DCis1c3} and
\eqref{DCis2c3} the following substitution should be used
\begin{eqnarray}\notag
f(\phi) \Delta^2&\to& \frac{r_0}{4 \pi
N}\zeta\left(2,\frac{1}{2}+\frac{1}{4\pi T_c \tau}\right )\left
[\gamma\left (\frac{1}{4\pi T_c \tau}\right )\right
]^{-1/2}  \\
&\times & \sqrt{\frac{T_c}{T-T_c}}\int \limits_{0}^{2\pi} \frac{d
\phi}{2\pi} \frac{f(\phi)}{\sqrt{1 + \displaystyle\frac{ \eta
T_c}{T-T_c}\sin^2\phi}}.\label{fl2c3}
\end{eqnarray}
It is worthwhile to mention that in order to obtain the result for
$T < T_c$ from the known result for $T>T_c$ we should substitute
$2 (T_c-T_c)/T_c$ for $(T-T_c)/T_c$ as usual.

\subsection{\label{ch3.3.1}Fluctuational correction to the anisotropic conductivity $\sigma_{ab}^{(\rm anis)}$}

\begin{figure}[t]
\centerline{\includegraphics[width=70mm]{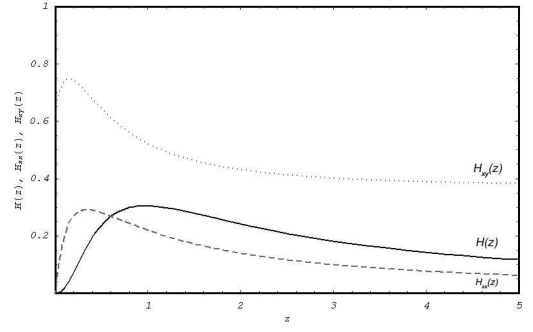}}
\caption{Functions $H(z)$, $H_{xx}(z)$ and $H_{xy}(z)$.}
\label{FIG4.4}
\end{figure}

Integrating over angle $\phi$ in Eq.\eqref{fl2c3} for
$f(\phi)=\cos 2\phi$, we obtain the following fluctuational
corrections to the anisotropic part of the conductivity tensor
above and below $T_c$
\begin{eqnarray}
\left . \begin{array}{c}
 \delta\sigma_{xx}^{(\rm anis-f)} \\
 \delta\sigma_{yy}^{(\rm anis-f)}
\end{array}
\!\!\right \} &=& \mp r_0 \mathcal{J}_1^2(r_0)H\left
(\frac{1}{4\pi T_c \tau}\right )\sqrt{\frac{T_c}{T-T_c}}\quad \notag\\
&\times & \begin{cases}
     \displaystyle\frac{1}{\sqrt{2}}F_A\left (\frac{\eta T_c}{2(T_c-T)}\right ),\!\! &\!\! T<T_c, \\
    \displaystyle F_A\left (\frac{\eta T_c}{T-T_c}\right ),\!\! &\!\! T>
    T_c.
  \end{cases}\label{flrc3a}
\end{eqnarray}
Here function $H(z)$ is determined by the function $h(z)$ as
\begin{equation}\label{flaHc3}
H(z) =
\frac{\zeta(2,\frac{1}{2}+z)h(z)}{\sqrt{\gamma(z)}}=\!\!\begin{cases}
    \displaystyle \frac{2\pi^4}{3\sqrt{\beta_1}} z^3, & z \ll 1, \\
    \displaystyle \frac{\sqrt{3}}{z\sqrt{\mathcal{J}_1^2(r_0)+3\beta_1}
    },
      & z \gg 1.
  \end{cases}
\end{equation}
We note that the function $h(z)$ has monotonic growth whereas
function $\zeta\left (2,\frac{1}{2}+z\right )/\sqrt{\gamma(z)}$
monotonically decreases. As a result the function $H(z)$ has
maximum at $z \approx 0.97$ (see Fig.~\ref{FIG4.4}). Function
$F_A(x)$ involves complete elliptic functions of the first and
second kind
\begin{eqnarray}\notag
F_A(x) &=& \frac{2}{\pi}\left [\left (1+ \frac{2}{x}\right )
K(i\sqrt{x}) - \frac{2}{x} E(i\sqrt{x})\right ] \\&=&
\begin{cases}
    \displaystyle \frac{x}{8}, & x \ll 1, \\
    \displaystyle \frac{\pi}{\sqrt{x}}\ln 16 e^{-4}x,
      & x \gg 1.
  \end{cases}\label{flaFAc3}
\end{eqnarray}

Integration over angle $\phi$ in Eq.\eqref{fl2c3} with
$f(\phi)=\sin 2\phi$ vanishes. Thus, fluctuational corrections to
the anisotropic part of $\sigma_{xy}$ and $\sigma_{yx}$
conductivities are absent if the UCDW is oriented at the angle
$\phi=0$ with respect to the $x$ axis. We mention that in this
case the mean-field contribution to $\sigma_{xy}$ and
$\sigma_{yx}$ vanishes as well. Consequently, the off-diagonal
components of the conductivity tensor are isotropic for $\phi=0$.

It is worthwhile to emphasize that Eq.\eqref{flrc3a} constitutes
one the main results of the present paper.

\subsection{\label{ch3.3.2}Fluctuational correction to the anisotropic conductivity  $\sigma_{ab}^{(\rm isot)}$}

Integrating over angle $\phi$ in Eq.\eqref{fl2c3} with
$f(\phi)=1$, we obtain the following fluctuational corrections to
the isotropic part of the conductivity tensor above and below
$T_c$
\begin{eqnarray}\label{flrc3}
\delta\sigma_{xx}^{(\rm isot-f)} &=& -r_0H_{xx}\left
(\frac{1}{4\pi T_c \tau}\right )\sqrt{\frac{T_c}{T-T_c}}\quad \\
&\times &
\begin{cases}
\displaystyle\frac{1}{\sqrt{2}} F_I\left (\frac{\eta T_c}{2(T_c-T)}\right ), & T<T_c, \\
    \displaystyle F_I\left (\frac{\eta T_c}{T-T_c}\right ),  & T> T_c,
\end{cases}\notag
\end{eqnarray}
and
\begin{eqnarray}\label{flr1c3}
 \delta\sigma_{xy}^{(\rm isot-f)} &=& -
\frac{2\pi r_0 T_c}{\omega_H} H_{xy}\left (\frac{1}{4\pi T_c
\tau}\right )\sqrt{\frac{T_c}{T-T_c}}\quad \\ &\times &
  \begin{cases}
\displaystyle\frac{1}{\sqrt{2}}F_I\left (\frac{\eta T_c}{2(T_c-T)}\right ), & T<T_c, \\
    \displaystyle F_I\left (\frac{\eta T_c}{T-T_c}\right ),  & T>
    T_c.
  \end{cases}\notag
\end{eqnarray}
Here functions $H_{xx}(z)$ and $H_{xy}(z)$ can be expressed via
functions $h_{xx}(z)$ and $h_{xy}(z)$ as follows
\begin{eqnarray}\label{flr2c3}
H_{xx}(z) &=& \frac{\zeta(2,\frac{1}{2}+z)
h_{xx}(z)}{\sqrt{\gamma(z)}} \\ &=&
  \begin{cases}
    \displaystyle \frac{\pi^2}{3\sqrt{\beta_1}} z, &\quad z \ll 1, \\
    \displaystyle \frac{\sqrt{3}(1+4 \mathcal{J}_1^2(r_0))}{4 \sqrt{\mathcal{J}_1^2(r_0)+3\beta_1}
    }\frac{1}{z},
      &\quad z \gg 1,
  \end{cases}\notag
\end{eqnarray}
\begin{eqnarray}\label{flr3c3}
H_{xy}(z) &=&
\frac{\zeta(2,\frac{1}{2}+z)h_{xy}(z)}{\sqrt{\gamma(z)}} \\ &=&
\begin{cases}
    \displaystyle \frac{1}{\sqrt{\beta_1}}\left (1 -
    \frac{\pi^2-2\psi^{''}(\frac{1}{2})}{\pi^2}z
    \right ), &\quad z \ll 1, \\
    \displaystyle \frac{3 \pi-8 \mathcal{J}_1^2(r_0)}{4 \sqrt{3}\sqrt{\mathcal{J}_1^2(r_0)+3\beta_1}
    },
      &\quad z \gg 1.
  \end{cases}\notag
\end{eqnarray}
We mention that functions $H_{xx}(z)$ and $H_{xy}(z)$ have maximum
at $z$ equal to $0.34$ and $0.16$, respectively, as it is shown in
Fig.~\ref{FIG4.4}. Function $F_I(x)$ involves complete elliptic
integral of the first kind
\begin{equation}\label{FI}
F_I(x) = \frac{2}{\pi}K(i\sqrt{x}) = \begin{cases}
    \displaystyle 1, &\quad x \ll 1, \\
    \displaystyle \frac{1}{\pi\sqrt{x}}\ln 16 x,
      &\quad x \gg 1.
  \end{cases}
\end{equation}

We emphasize that Eqs.\eqref{flr2c3}-\eqref{flr3c3} are one of the
main results of the present paper.

\subsection{\label{ch3.3.3} Limit of applicability of Eqs.\eqref{flrc3a}, \eqref{flr2c3} and \eqref{flr3c3}}

Eqs.\eqref{flrc3a}, \eqref{flr1c3} and \eqref{flr2c3} have
singularity at $T\to T_c$. It indicates that the results are not
applicable near $T_c$. The limit of their applicability is
determined by the condition that fluctuational corrections
\eqref{flrc3a}, \eqref{flr1c3} and \eqref{flr2c3} are still small
as compared to conductivity of the liquid state equal
to~\cite{AFS}
\begin{equation}\label{hets}
\sigma_{xx}^{(0)} = \frac{2 N}{\pi}, \qquad \sigma_{xy}^{(0)} = N.
\end{equation}

Below we prove that the condition of smallness of fluctuational
corrections and the condition $|T_c-T|\ll T_c$ are compatible. Let
us first consider Eq.\eqref{flr1c3}. By using the facts that $\max
F_I(x)=1$ and $\max H_{xx}(z)\approx 0.3$ (see Fig.~\ref{FIG4.4})
and Eq.\eqref{hets}, we obtain
\begin{equation}\label{het2}
1 \gg \frac{|T_c-T|}{T_c} \gg N^{-2}.
\end{equation}
As one can see, Eq.\eqref{het2} is fulfilled in the limit $N\gg
1$. Analysis of Eq.\eqref{flrc3a} results in similar non-equality
with $0.1$ instead of $1$ in the right hand side.
Eq.\eqref{flr2c3} contains additional small factor $T_c/\omega_H$.
Therefore, the condition of applicability for results
\eqref{flrc3a}, \eqref{flr1c3} and \eqref{flr2c3} is given by
Eq.\eqref{het2}.

%
\section{\label{ch3.4}Discussion}

\begin{figure}[t]
\centerline{\includegraphics[width=70mm]{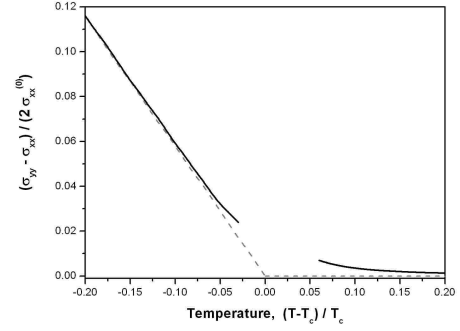}}
\caption{The dependence of anisotropic part of conductivity
$(\sigma_{yy}-\sigma_{xx})/2\sigma_{xx}^{(0)}$ on temperature for
$1/4\pi T_c\tau = 0.24$, $\eta=0.01$ and $N=2$. Dashed line
corresponds to mean-field result \eqref{DCanis1c3}.}
\label{FIG4.5}
\end{figure}

In the previous sections we derived a number of results for the
conductivity tensor of two-dimensional interacting electrons on a
half-filled high Landau level. We demonstrated that below
temperature $T_c$ of transition from the liquid to the UCDW state
the anisotropic part of the conductivity tensor emerges. At
$(T_c-T)/T_c \ll 1$ the anisotropic part is proportional to
deviation of temperature from $T_c$. As it is shown in
Fig.~\ref{FIG4.5}, it results in a cusp of temperature dependence
of the conductivity at $T=T_c$. Fluctuations of the order
parameter above and below transition temperature $T_c$ smooth out
the cusp (as it is shown in Fig.~\ref{FIG4.5}).

Results discussed above have been derived in the case of the
white-noise random potential. In the case of a random potential
with correlation length $d$ arbitrary related with the magnetic
length $l_H$, we can state that temperature dependence of the
results above remain the same whereas the functions $h_{ab}$ and
$H_{ab}$ become to depend not only on $1/4\pi T_c \tau$ but on
ratio $d/l_H$ also.

In experiments~\cite{LCEPW,DTSPW} magnetoresistance $R_{xx}$ and
$R_{yy}$ as functions of temperature at filling factor $\nu=9/2$
have been investigated. Unfortunately, the detailed temperature
dependence near the point at which $R_{xx}$ and $R_{yy}$ become to
deviate from each other did not investigated. Nevertheless,
results reported in Refs.~\cite{LCEPW,DTSPW} confirm the linear
dependence of magnetoresistance $R_{xx}$ and $R_{yy}$ on
temperature in the certain range of temperatures not too close to
the $T_c$. Therefore, the detailed investigation of temperature
dependence of magnetoresistance are needed in the future.

The results \eqref{DCanis1c3} and \eqref{DCanis2c3} for the angle
dependence of the anisotropic part of the conductivity tensor
qualitatively describe the results of
experiments~\cite{LCEPW,DTSPW,ELCPW}. If a current runs in the
direction of charge density modulation, the conductivity
(resistance) is less (more) than the conductivity (resistance) for
the case of current running in the direction perpendicular to
charge density modulation. The Hall conductivity (resistance)
remains roughly the same.

The UCDW state exists also near half-filled high Landau level. In
the case of $\mu_N\neq 0$, the functions $h_{ab}$ and $H_{ab}$ are
dependent not only on parameter $1/4\pi T_c\tau$ but on ratio
$\mu_N/T_c$ also. Increasing the chemical potential $\mu_N$ we
transform the UCDW state to the triangular CDW state characterized
by vectors $\textbf{Q}_1$, $\textbf{Q}_2$ and $\textbf{Q}_3$
directed at angles $\phi$, $\phi+2\pi/3$ and $\phi+4\pi/3$ with
respect to the $x$ axis, respectively. However, due to the
identity
\begin{equation}\label{disc1ch3}
1 + e^{2\pi i/3} +e^{-2\pi i/3} = 0
\end{equation}
the contribution from three vectors $\textbf{Q}_i$ to the
anisotropic part of conductivity tensor vanishes. Thus, the
conductivity tensor of the triangular CDW state remains isotropic
in the approximation which is of the second order in the CDW order
parameter. From general physical arguments it is clear that the
conductivity tensor of the triangular CDW state should be
isotropic.

It is worthwhile to mention that from physical point of view the
anisotropic part $\sigma_{ab}^{(\rm anis)}$ of conductivity
appears due to the existence of the induced anisotropic potential
$\lambda(\textbf{r})$ in action \eqref{SHF5c3}. Anisotropic
resistance of two-dimensional electrons in a weak magnetic field
in the presence of unidirectional periodic potential has been
measured at several Kelvins in heterostructures with mobility
$\mu_0\sim 10^6\, cm^2/V s$ by Weiss, von Klitzing, Ploog, and
Weimann fifteen years ago~\cite{WKPW}. Theoretically, the effect
of unidirectional periodic potential on the conductivity tensor of
two-dimensional electrons in a weak magnetic field has been
investigated with the help of both the kinetic
equation~\cite{Bee,MW} and the diagrammatic
technique~\cite{GWK,ZG,PV}. However, the case considered in the
present paper did not analyze in the
Refs.~\cite{Bee,MW,GWK,ZG,PV}. The important difference of the
induced potential $\lambda(\textbf{r})$ from external periodic
potential is that the induced potential $\lambda(\textbf{r})$
appears on the $N$th Landau level only whereas external potential
scatters electrons on all Landau levels.

%
\section{\label{ch3.5}Conclusion}

We obtained the conductivity tensor of two-dimensional electrons
in the presence of weak disorder and weak magnetic field at
half-filled high Landau level where the UCDW state exists. In the
framework of the order parameter expansion we derived that at
$T_c-T\ll T_c$ anisotropic part of the conductivity tensor
proportional to $(T_c-T)/T_c$ emerges. Also we demonstrated that
the order parameter fluctuations result in additional anisotropic
contributions near $T_c$ to the conductivity tensor that wash out
the mean-field cusp at $T=T_c$. The results obtained are in
agreement with the experimental findings.

\acknowledgements  I am grateful to M.A. Baranov, L.I. Glazman,
M.V. Feigelman, S.V. Iordansky, P.M. Ostrovsky, M.A. Skvortsov for
discussions and especially to R.R. Du for experimental plots.
Financial support from Russian Foundation for Basic Research
(\textit{RFBR}), the Russian Ministry of Science, Dynasty
Foundation, Forschungszentrum J\"ulich (\textit{Landau
Scholarship}), and Dutch Science Foundation (\textit{FOM}) is
acknowledged.

\appendix


%
\section{\label{ch3.App.1}Calculation of characteristic energy $T_1$}

According to Eqs.\eqref{UHFc3} and \eqref{T0T1c3} the $T_1(q)$ is
given by
\begin{eqnarray}\label{app1c3}
T_1(q) &=& - n_L e^{i\phi} \Biggl (U_{\rm scr}(q) \mathcal{J}_0(q
R_c) \mathcal{J}_1(q R_c) e^{-i\phi} \\ &-& \sqrt{2N} \int \frac{d
\textit{\textbf{p}}}{(2\pi)^2 n_L} e^{-i \textbf{q}
\textit{\textbf{p}} l_H^2} U_{\rm scr}(p)   \frac{p_x-ip_y}{p^2
l_H} \notag\\
&\times &L_{NN}\left (\frac{p^2 l_H^2}{2}\right ) L_{N,N-1}\left
(\frac{p^2 l_H^2}{2}\right ) e^{-p^2 l_H^2/2} \Biggr ),\notag
\end{eqnarray}
where we use the following result~\cite{AFS}
\begin{eqnarray}\notag
F_{N,N-1}(\textbf{q}) &=& \sqrt{2N} \frac{q_x-i q_y}{q^2 l_H}
e^{-q^2 l_H^2/4}  L_{N,N-1}\left (\frac{q^2 l_H^2}{2}\right )\\ &
\approx & e^{-i \phi} \mathcal{J}_1(q R_c), \,\, q R_c \ll
2N.\label{app2c3}
\end{eqnarray}
The characteristic energy $T_1 = T_1(Q_0)$ is given by
\begin{equation}\label{app3c3}
T_1 = \frac{r_s \omega_H}{4 \sqrt{2}} \int \limits_{0}^{4N}
\frac{d x}{\tilde\epsilon(x)} \frac{\mathcal{J}_1(4 N
x)\mathcal{J}_0(4N x)}{\sqrt{1-x^2}} \mathcal{J}_1(2 r_0 x),
\end{equation}
where
\begin{equation}\label{app4c3}
\tilde \epsilon(x) = 1 + \frac{r_s}{x\sqrt{2}}(1-\mathcal{J}_0^2(4
N x)).
\end{equation}
Performing calculation of the integral, we find
\begin{equation}\label{app5c3}
T_1= \frac{r_s\omega_H}{16 \pi N \sqrt{2}} \left [\frac{r_0}{2}
\ln \left ( 1+\frac{1}{\sqrt{2} r_0 r_s}\right ) +
\frac{c_1}{1+\sqrt{2} r_0 r_s}\right ],
\end{equation}
where constant $c_1$ equals
\begin{equation}\label{app6c3}
c_1 = \sqrt{\frac{r_0}{\pi}} \int \limits_{1/2r_0}^1 \frac{d x }{x
\sqrt{x(1-x^2)}} \sin \left (2 r_0 x- \frac{\pi}{4}\right )\approx
1.097.
\end{equation}
As we can see from Eq.\eqref{app5c3} the characteristic energy
$T_1\sim T_0/N$ as we mentioned above.

%
\section{\label{ch3.App.2}Calculation of the $I_{p_1p_2p_3p_4}(\textbf{Q}_0)$}

Using definition \eqref{Ic3} and Eq.\eqref{app2c3}, we obtain in
the limit $N\gg 1$
\begin{eqnarray}\notag
I_{N,N-1,N-1,N} &=& I_{N,N+1,N+1,N} = g n_L N \int
\limits_{0}^{\infty} \frac{d x}{x} e^{-x} \\ &\times & \left
[L_{N}^{-1}(x)\right ]^2
\mathcal{J}_0(r_0\sqrt{2x}),\label{app7c3} \\
I_{N,N-1,N,N+1} &=& I_{N,N+1,N,N-1} = g n_L N \int
\limits_{0}^{\infty} \frac{d x}{x} e^{-x} \notag \\ &\times &
L_{N}^{-1}(x) L_{N+1}^{-1}(x)
\mathcal{J}_0(r_0\sqrt{2x}),\notag\\
I_{N,N,N,N-1}  &=& -I_{N,N,N,N+1} = g n_L e^{i\phi} \sqrt{N} \int
\limits_{0}^{\infty} \frac{d x}{\sqrt{x}} e^{-x} \notag \\
&\times & L_{N}^{-1}(x) L_{N}(x)
\mathcal{J}_1(r_0\sqrt{2x}),\notag\\
I_{N,N+1,N,N}  &=& -I_{N,N+1,N,N} = g n_L e^{-i\phi} \sqrt{N} \int
\limits_{0}^{\infty} \frac{d x}{\sqrt{x}}  \notag \\
&\times & e^{-x} L_{N+1}^{-1}(x) L_{N}(x)
\mathcal{J}_1(r_0\sqrt{2x}),\notag\\
I_{N,N-1,N,N-1} &=& I^{*}_{N,N+1,N,N+1} = g n_L e^{2 i \phi} N
\int \limits_{0}^{\infty} \frac{d x}{x} e^{-x} \notag \\
&\times & \left [L_{N}^{-1}(x)\right ]^2
\mathcal{J}_2(r_0\sqrt{2x}), \notag \\
I_{N,N-1,N+1,N} &=& I^{*}_{N,N+1,N-1,N} = g n_L e^{2 i \phi} N
\int \limits_{0}^{\infty} \frac{d x}{x} e^{-x} \notag \\ &\times &
L_{N}^{-1}(x) L_{N+1}^{-1}(x) \mathcal{J}_2(r_0\sqrt{2x}).\notag
\end{eqnarray}
With a help of asymptotic expression for Laguerre
polynomial~\cite{GR}
\begin{gather}
L_N^\alpha(x) \simeq \frac{1}{\sqrt{\pi x}} e^{x/2} \left
(\frac{n}{x}\right )^{(2\alpha-1)/4} \cos \left (2\sqrt{n x}
-\frac{\alpha\pi}{2}-\frac{\pi}{4} \right ),\notag\\ N \gg
1,\label{app8c3}
\end{gather}
we find
\begin{eqnarray}\label{app9c3}
I_{N,N-1,N-1,N} &=& I_{N,N+1,N+1,N} \\ &=& g n_L \frac{2}{\pi}
\int \limits_{0}^{1} d x \frac{\mathcal{J}_0(2 r_0
x)}{\sqrt{1-x^2}}, \notag
\end{eqnarray}
\begin{eqnarray}
 I_{N,N-1,N,N+1} &=& I_{N,N+1,N,N-1}
\notag \\ &=& g n_L \frac{2}{\pi} \int \limits_{0}^{1} d x (1-2
x^2) \frac{\mathcal{J}_0(2 r_0 x)}{\sqrt{1-x^2}},\notag
\end{eqnarray}
\begin{eqnarray}
I_{N,N,N,N-1}  &=& -I_{N,N,N,N+1} \notag \\ &=& g n_L e^{i\phi}
\frac{2}{\pi} \int \limits_{0}^{1} d x x \frac{\mathcal{J}_1(2 r_0
x)}{\sqrt{1-x^2}},\notag
\end{eqnarray}
\begin{eqnarray}
I_{N,N+1,N,N}  &=& -I_{N,N+1,N,N} \notag \\ &=& g n_L e^{-i\phi}
\frac{2}{\pi} \int \limits_{0}^{1} d x x \frac{\mathcal{J}_1(2 r_0
x)}{\sqrt{1-x^2}},\notag
\end{eqnarray}
\begin{eqnarray}
I_{N,N-1,N,N-1} &=& I^{*}_{N,N+1,N,N+1} \notag \\ &=& g n_L e^{2 i
\phi} \frac{2}{\pi} \int \limits_{0}^{1} d x \frac{\mathcal{J}_2(2
r_0 x)}{\sqrt{1-x^2}}, \notag
\end{eqnarray}
\begin{eqnarray}
I_{N,N-1,N+1,N} &=& I^{*}_{N,N+1,N-1,N} \notag \\ &=& g n_L e^{2 i
\phi} \frac{2}{\pi}  \int \limits_{0}^{1} d x (1-2 x^2)
\frac{\mathcal{J}_2(2 r_0 x)}{\sqrt{1-x^2}}.\notag
\end{eqnarray}
The integrals can be evaluated by using the following
equality~\cite{GR}
\begin{equation}\label{app10c3}
\frac{2}{\pi} \int\limits_{0}^{\pi/2} d \phi \cos(2\mu \phi)
\mathcal{J}_{2\nu}(2 r_0 \cos \phi) =
\mathcal{J}_{\nu+\mu}(r_0)\mathcal{J}_{\nu-\mu}(r_0).
\end{equation}
Finally, it yields the results presented in Table~\ref{Tablec31}.



\begin{thebibliography}{99}
\bibitem{Foot1} In the clean case ($\tau^{-1}\to 0$) the screened
electron-electron interaction was found in Ref.~\cite{AG}.
\bibitem{Foot2} In the approximation quadratic in the order parameter $\Delta_{p_1p_2}(\textbf{Q})$
the contributions to the thermodynamic potential and to the free
energy are the same.
\bibitem{Foot3} We use the following assymptotic expression for Riemann zeta-function
$\zeta(k,z) = (k-1)^{-1}z^{1-k} - 2^{-1} z^{-k}$ at $z \gg 1$ and
the fact that $\zeta(6,1/2) = \pi^6/15$.
\end{thebibliography}
\end{document}